\documentclass[12pt]{article}
\newcommand{\be}{\begin{eqnarray}}
\newcommand{\ee}{\end{eqnarray}}
\newcommand{\non}{\nonumber}
\newcommand{\tr}{\mathop{\rm tr}\nolimits}

\begin{document}

\begin{titlepage}
\strut\hfill UMTG--215
\vspace{.5in}
\begin{center}

\LARGE Parity and Charge Conjugation Symmetries\\
\LARGE and $S$ Matrix of the XXZ Chain \\[1.0in]
\large Anastasia Doikou and Rafael I. Nepomechie\\[0.8in]
\large Physics Department, P.O. Box 248046, University of Miami\\[0.2in]  
\large Coral Gables, FL 33124 USA\\

\end{center}

\vspace{.5in}

\begin{abstract}
We formulate the notion of parity for the periodic XXZ spin chain 
within the Quantum Inverse Scattering Method.  We also propose an 
expression for the eigenvalues of the charge conjugation operator.  We 
use these discrete symmetries to help classify low-lying $S^{z}=0$ 
states in the critical regime, and we give a direct computation of 
the $S$ matrix.
\end{abstract}

\end{titlepage}

\section{Introduction and summary}

The periodic anisotropic Heisenberg (or ``XXZ'' ) spin chain, with the 
Hamiltonian
\be
H =  {1\over 4} \sum_{n=1}^{N} \left\{
\sigma^{x}_n \sigma^{x}_{n+1}
+ \sigma^{y}_n \sigma^{y}_{n+1} 
+ \Delta \left( \sigma^{z}_n \sigma^{z}_{n+1} - 1 \right) \right\} \,, 
\qquad \vec \sigma_{N+1} \equiv \vec \sigma_{1} \,, 
\label{hamiltonian} 
\ee 
has a long and rich history \cite{bethe} - \cite{kbi}.  It is 
the prototype of all integrable models.  The development of the 
Quantum Inverse Scattering Method (QISM)/algebraic Bethe Ansatz 
\cite{faddeev/takhtajan} systematized earlier results, and paved the 
way for far-reaching generalizations.

Parity symmetry has played a valuable role in continuum quantum field 
theory, including integrable quantum field theory.  (See, e.g., 
\cite{korepin}.) However, the notion of parity for discrete spin 
models, in particular for those which are integrable, has not (to our 
knowledge) been discussed.  We show here that parity has a simple 
realization in the algebraic Bethe Ansatz, involving negation of the 
spectral parameter, i.e., $\lambda \rightarrow -\lambda$.  (See Eq.  
(\ref{parity/second}) below.)

We consider also charge conjugation symmetry \cite{baxter}.  We 
conjecture that Bethe Ansatz states of the XXZ chain with $S^{z}=0$ 
are eigenstates of the charge conjugation operator, with eigenvalues 
$(-1)^{\nu}$, where ${\nu}$ is given by Eq.  (\ref{conjecture}).

Working within the framework of the string hypothesis 
\cite{takahashi/suzuki}, we use these discrete symmetries to help 
classify low-lying $S^{z}=0$ states \cite{jkm}, \cite{woynarovich} in 
the critical regime with $0 < \Delta < 1$.  Moreover, we compute the 
$S$ matrix elements corresponding to these states using the method of 
Korepin \cite{korepin} and Andrei-Destri \cite{andrei/destri}.  Our 
results for the $S$ matrix agree with those obtained by thermodynamic 
methods \cite{babujian/tsvelick}, \cite{kirillov/reshetikhin}.

The outline of this paper is as follows.  In Section 2, after a brief 
review of the algebraic Bethe Ansatz, we define the parity operator, 
and we show how it acts on the fundamental quantities of the QISM 
formalism.  We also review the definition of the charge conjugation 
operator, and we propose an expression for the corresponding 
eigenvalues.  In Section 3 we compute root densities for low-lying 
states in the critical regime.  We use these densities to calculate 
quantum numbers -- in particular, parity and charge conjugation -- of 
the $S^{z}=0$ states, as well as the $S$ matrix.  We conclude in 
Section 4 by briefly listing some remaining unanswered questions.
This paper is an expanded version of a recent letter \cite{d/n}.

\section{Algebraic Bethe Ansatz and discrete symmetries}

In order to fix notations, we briefly recall the essential elements of 
the algebraic Bethe Ansatz for the XXZ chain.  (See the above-cited 
references for details.) We consider the $A^{(1)}_{1}$ $R$ matrix
\be
R(\lambda)=\left( \begin{array}{cccc}
	a(\lambda)                                  \\
    &         b(\lambda) & c                    \\
	&         c          & b(\lambda)           \\
	&                    &           & a(\lambda)
\end{array} \right) \,, 
\ee 
where
\be
a(\lambda) &=& {\sinh \left( \mu (\lambda + i) \right)\over \sinh (i \mu)} 
\non  \\ 
b(\lambda) &=& {\sinh (\mu \lambda)\over \sinh (i \mu)} \non  \\
c &=& 1 \,.
\ee
We regard $R(\lambda)$ as an operator acting on the tensor product 
space $V \otimes V$, where $V$ is a two-dimensional complex vector 
space. This $R$ matrix is a solution of the Yang-Baxter equation
\be
R_{12}(\lambda-\lambda')\ R_{13}(\lambda)\ R_{23}(\lambda')
= R_{23}(\lambda')\ R_{13}(\lambda)\ R_{12}(\lambda-\lambda') \,,
\label{YB}
\ee 
where $R_{ij}$ are operators on $V \otimes V \otimes V$, with $R_{12} 
= R \otimes 1$, etc.  We define the $L$ operators \footnote{We make the 
shift in the spectral parameter in order for the Bethe Ansatz 
equations to have the standard symmetric form given in Eq.  
(\ref{BAE}) below.}
\be
L_{0 n}(\lambda) = R_{0 n}(\lambda - {i\over 2}) 
= \left( \begin{array}{cc}
\alpha_{n}(\lambda) & \beta_{n}  \\
\gamma_{n}          & \delta_{n}(\lambda)
\end{array} \right) 
\,,
\ee  
which act on so-called auxiliary ($0$) and quantum ($n$) spaces.  
Evidently,
\be
\alpha(\lambda) &=& \left(\begin{array}{cc}
                a(\lambda-{i\over 2}) & 0 \\
				0 & b(\lambda-{i\over 2})
			\end{array} \right) \,, \qquad 
			\beta = \left(\begin{array}{cc}
                0 & 0 \\
				c & 0
			\end{array} \right) \,, \non \\
\gamma &=& \left(\begin{array}{cc}
                0 & c \\
                0 & 0
			\end{array} \right) \,, \qquad 
			\delta(\lambda)  = \left(\begin{array}{cc}
               b(\lambda-{i\over 2}) & 0 \\
			   0 & a(\lambda-{i\over 2})
			\end{array} \right) \,.			
\ee 
The monodromy matrix $T_{0}(\lambda)$ is defined as a product of $N$ 
such operators
\be
T_{0}(\lambda) = L_{0 N}(\lambda) \cdots L_{0 1}(\lambda) =
\left( \begin{array}{cc}
A(\lambda) & B(\lambda)  \\
C(\lambda) & D(\lambda)
\end{array} \right) 
\,.
\ee
(As is customary, we often suppress the quantum-space subscripts.)
It obeys the fundamental relation
\be
R_{12}(\lambda-\lambda')\ T_{1}(\lambda)\ T_{2}(\lambda')
= T_{2}(\lambda')\ T_{1}(\lambda)\ R_{12}(\lambda-\lambda') \,.
\label{fundamental}
\ee 
The transfer matrix $t(\lambda)$, defined by tracing over the 
auxiliary space
\be
t(\lambda) = \tr_{0} T_{0}(\lambda) = A(\lambda) + D(\lambda) \,,
\ee
has the commutativity property
\be
\left[ t(\lambda)\,, t(\lambda') \right] = 0 
\ee
by virtue of the fundamental relation (\ref{fundamental}).  The 
transfer matrix also commutes with the $z$ component of the total 
spin,
\be
\left[ t(\lambda)\,, S^{z} \right] = 0 \,, \qquad 
S^{z} = {1\over 2} \sum_{n=1}^{N} \sigma^{z}_n \,.
\label{spin}
\ee 
The Hamiltonian
\be
H &=& {i \sin \mu \over 2 \mu} {d\over d \lambda} \log 
t(\lambda)\Big\vert_{\lambda = {i\over 2}} - {N\over 2}\cos \mu  \non \\
&=& \sum_{n=1}^{N-1} H_{n\,, n+1} + H_{N \,,1} \,, \qquad 
H_{ij} = {i \sin \mu \over 2\mu} {\cal P}_{ij}\ R_{ij}'(0) 
- {1\over 2}\cos \mu 
\label{twosite}
\ee
coincides with the XXZ Hamiltonian (\ref{hamiltonian}), provided
\be
\Delta = \cos \mu \,.
\ee 
The critical regime $-1 < \Delta < 1$ corresponds to $\mu$ real, with 
$0 < \mu  < \pi$. The momentum operator $P$ is defined by
\be
P={1\over i}\log \ t({i\over 2})  \,,
\label{momentum}
\ee
since $t({i\over 2}) $ is the one-site shift operator.

Let $\omega_{+}$ be the ferromagnetic vacuum vector with all spins up,
\be
\omega_{+} =  \underbrace{{1 \choose 0}
\otimes \cdots \otimes {1 \choose 0}}_{N} \,,
\ee
which is annihilated by $C(\lambda)$. The operators $B(\lambda)$,
which commute among themselves,
\be
\left[ B(\lambda) \,, B(\lambda') \right] = 0 \,,
\ee
act as creation operators. The Bethe state
\be
B(\lambda_{1}) \cdots B(\lambda_{M})\ \omega_{+}
\label{vec}
\ee
is an eigenstate of the transfer matrix $t(\lambda)$, with eigenvalue
\be
\Lambda(\lambda) &=& \left( {\sinh \mu ( \lambda + {i\over 2} )
\over \sinh \mu i} \right)^{N}
\prod_{\alpha=1}^{M} 
{\sinh  \mu \left( \lambda - \lambda_{\alpha} - i \right) 
\over 
 \sinh  \mu \left( \lambda - \lambda_{\alpha} \right) } \non  \\
& \ & + \left( {\sinh \mu ( \lambda - {i\over 2} )
\over \sinh \mu i} \right)^{N}
\prod_{\alpha=1}^{M} 
{\sinh  \mu \left( \lambda - \lambda_{\alpha} + i \right) 
\over 
 \sinh  \mu \left( \lambda - \lambda_{\alpha} \right) } 
\,,
\ee
if $\{ \lambda_{1} \,, \ldots \,, \lambda_{M} \}$ are distinct and obey 
the Bethe Ansatz equations
\be
\left( {\sinh  \mu \left( \lambda_{\alpha} + {i\over 2} \right) 
\over   \sinh  \mu \left( \lambda_{\alpha} - {i\over 2} \right) } 
\right)^{N} 
= \prod_{\scriptstyle{\beta=1}\atop \scriptstyle{\beta \ne \alpha}}^M 
{\sinh  \mu \left( \lambda_{\alpha} - \lambda_{\beta} + i \right) 
\over 
 \sinh  \mu \left( \lambda_{\alpha} - \lambda_{\beta} - i \right) }
\,, \qquad \alpha = 1 \,, \cdots \,, M \,. 
\label{BAE}
\ee
In particular, the energy and momentum are given by
\be
E &=& - \sin^{2} \mu  \sum_{\alpha=1}^{M} 
{1\over \cosh (2 \mu \lambda_{\alpha}) - \cos \mu } \,, \non \\ 
P &=& {1\over i} \sum_{\alpha=1}^{M} 
\log {\sinh  \mu \left( \lambda_{\alpha} + {i\over 2} \right) 
\over \sinh  \mu \left( \lambda_{\alpha} - {i\over 2} \right)} 
\quad (\mbox{mod } 2 \pi)
\,.
\ee 
The vector (\ref{vec}) is also an eigenvector of $S^{z}$ with 
eigenvalue 
\be 
S^{z}={N\over 2} - M \,.
\label{spineigenvalue}
\ee 

Having reviewed the QISM description of the XXZ model, we now consider 
some of its discrete symmetries.

\subsection{Parity}

We define the parity operator $\Pi$ on a ring of $N$ spins by
\be
\Pi \ X_{n} \ \Pi^{-1} = X_{N+1-n} \,,
\ee 
where $X_{n}$ is any operator at site $n \in \{ 1 \,, 2\,, \ldots \,, 
N \}$.  Clearly, $\Pi$ acts on the tensor product space $V^{\otimes 
N}$.  We can represent $\Pi$ by
\be
\Pi = \left\{ \begin{array}{cc}
{\cal P}_{1\,, N} {\cal P}_{2\,, N-1} \ldots {\cal P}_{{N\over 2}\,, 
{N+2\over 2}} & \quad \mbox{for} \quad N=\mbox{  even  } \\
{\cal P}_{1\,, N} {\cal P}_{2\,, N-1} \ldots {\cal P}_{{N-1\over 2}\,, 
{N+3\over 2}} & \quad \mbox{for} \quad N=\mbox{  odd  } 
\end{array} \right. \,,
\ee
where ${\cal P}_{i j}$ is the permutation matrix which permutes 
the $i^{th}$ and $j^{th}$ vector spaces. We note that 
$\Pi = \Pi^{-1} = \Pi^{\dagger}$ and hence $\Pi\ \Pi^{\dagger}=1$.

It is easy to see that both the Hamiltonian (\ref{hamiltonian}) and 
the spin (\ref{spin}) are invariant under parity, while the momentum 
(\ref{momentum}) changes sign
\be
\Pi \  H \ \Pi = H \,, \qquad \Pi \  S^{z} \ \Pi = S^{z} \,, \qquad 
\Pi \  P \ \Pi = - P \,.
\label{threeops}
\ee 
Indeed, the ``parity invariance'' of the $R$ matrix
\be
{\cal P}_{12}\ R_{12}(\lambda)\ {\cal P}_{12} =  R_{12}(\lambda)
\ee
implies that the two-site Hamiltonian (\ref{twosite}) satisfies
\be
{\cal P}_{ij}\ H_{ij}\ {\cal P}_{ij} =  H_{ij}
\,,
\ee
and hence the full Hamiltonian is parity invariant. Moreover, for any 
$X_{n}$, we have the chain of equalities
\be
\Pi \ t({i\over 2})\ X_{n}\ t({i\over 2})^{-1}\ \Pi = \Pi\ X_{n+1}\ \Pi 
= X_{N-n} = \non \\
= t({i\over 2})^{-1}\ X_{N+1-n}\ t({i\over 2}) =
t({i\over 2})^{-1}\ \Pi\ X_{n}\ \Pi\ t({i\over 2}) \,,
\ee
which is consistent with $\Pi\ t({i\over 2})\ \Pi = t({i\over 2})^{-1} 
$, implying the third relation in Eq.  (\ref{threeops}).

In order to go further, we must investigate the behavior of the 
monodromy matrix under parity. To this end, we first note the 
``time-reversal'' invariance of the $R$ matrix
\be
R_{12}(\lambda)^{t_{1} t_{2}} = R_{12}(\lambda) \,,
\ee
where $t_{j}$ denotes transposition in the $j^{th}$ space, which
implies the identity 
\be
L_{0 n}(\lambda)^{t_{0}} = L_{0 n}(\lambda)^{t_{n}} \,;
\label{identity1}
\ee
and we observe that
\be
\alpha_{n}(-\lambda) = - \delta_{n}(\lambda) \,,
\ee 
from which it follows that
\be
L_{0 n}(\lambda)^{t_{n}} = W_{0}\ L_{0 n}(-\lambda)\ W_{0} \,,
\qquad  W = \left(\begin{array}{cc}
	               0  & 1 \\
				  -1  & 0 
\end{array} \right) \,.
\label{identity2}
\ee
We next observe that
\be
T_{0}(\lambda)^{t_{0}} &=& L_{0 1}(\lambda)^{t_{0}} \cdots  
L_{0 N}(\lambda)^{t_{0}} \non  \\ 
&=& L_{0 1}(\lambda)^{t_{1}} \cdots  
L_{0 N}(\lambda)^{t_{N}} \non  \\ 
&=& W_{0} L_{0 1}(-\lambda) W_{0} \cdots  
W_{0} L_{0 N}(-\lambda) W_{0} \non  \\ 
&=& (-)^{N-1} W_{0} L_{0 1}(-\lambda) \cdots  
L_{0 N}(-\lambda) W_{0} \,,
\ee
where the second line uses (\ref{identity1}), 
the third line uses (\ref{identity2}),
and the last line uses $W^{2} = -1$. Finally, we obtain
the desired result
\be
\Pi\ T_{0}(\lambda)\ \Pi &=& \Pi L_{0 N}(\lambda) \Pi \cdots 
\Pi L_{0 1}(\lambda) \Pi \non  \\ 
&=& L_{0 1}(\lambda) \cdots L_{0 N}(\lambda)  \non  \\ 
&=& (-)^{N-1} W_{0}\  T_{0}(-\lambda)^{t_{0}}\  W_{0} \,.
\label{result}
\ee
Note that under parity, the order of the $L$ operators in the 
monodromy matrix is reversed.

In particular, we see that
\be
\Pi\ t(\lambda)\ \Pi = (-)^{N} t(-\lambda) \,.
\label{parity/first}
\ee
Evidently, the parity operator does not commute with the transfer 
matrix.  Nevertheless, if $|v \rangle $ is an eigenvector of 
$t(\lambda)$ with eigenvalue $\Lambda(\lambda)$, then $\Pi |v \rangle $ 
is also an eigenvector of $t(\lambda)$, but with eigenvalue $(-)^{N} 
\Lambda(-\lambda)$.

We also obtain from (\ref{result}) the fundamental result
\be
\Pi\ B(\lambda)\ \Pi = (-)^{N-1} B(-\lambda) \,.
\label{parity/second}
\ee
We shall use this result, together with the fact
\be
\Pi\ \omega_{+} =  \omega_{+} \,,
\ee 
to investigate whether the eigenvectors (\ref{vec}) of the transfer 
matrix are also eigenvectors of the parity operator.  Eqs.  
(\ref{threeops}), (\ref{parity/first}) and (\ref{parity/second}) are 
the key results of this subsection.

We conclude this subsection with a few general observations.  We first 
note that if the set $\{ \lambda_{\alpha} \}$ is a solution of the 
Bethe Ansatz equations, then so is the corresponding set $\{ 
-\lambda_{\alpha} \}$ with each root negated.  As always, we assume 
that the Bethe Ansatz roots are distinct.  Moreover, for simplicity, 
let us assume (in this paragraph) that $N$ is even, and that the roots 
are nonzero and are not equal to ${i\pi\over 2 \mu}$.  For $M=$ even, 
all the roots can be ``paired'' $\{ \lambda_{\alpha} \}= \{ 
\lambda_{1} \,, -\lambda_{1} \,,\ldots \,, \lambda_{{M\over 2}} \,, 
-\lambda_{{M\over 2}} \}$, in which case $\{ -\lambda_{\alpha} \}$ is 
the same set. Eq.  (\ref{parity/second}) then implies that the 
corresponding state $\prod_{\alpha=1}^{M} B(\lambda_{\alpha}) 
\omega_{+}$ is a parity eigenstate, with eigenvalue $(-)^{M}$.  
Similarly, a nondegenerate state must have parity $(-)^{M}$.

There can be Bethe states which are not parity eigenstates. For such 
cases, the vectors $\prod_{\alpha} B(\lambda_{\alpha}) \omega_{+}$ and
$\prod_{\alpha} B(-\lambda_{\alpha}) \omega_{+}$ are distinct and are
degenerate in energy.

Since $\{ \Pi \,, P \} = 0$, a Bethe state can be an eigenstate of 
$\Pi$ only if the momentum is $P = 0 \mbox{ or } \pi \quad (\mbox{mod 
} 2 \pi)$.

\subsection{Quasi-periodicity}

The transfer matrix $t(\lambda)$ and the operators $B(\lambda)$ also 
satisfy quasi-periodicity conditions, which we now derive.  The 
starting point is the observation
\be
\alpha_{n}(\lambda \pm {i\pi\over \mu}) &=& - \alpha_{n}(\lambda) \,,
\non \\ 
\delta_{n}(\lambda \pm {i\pi\over \mu}) &=& - \delta_{n}(\lambda) \,,
\ee 
from which it follows that
\be
L_{0 n}(\lambda \pm {i\pi\over \mu}) = - S_{0}\ L_{0 n}(\lambda)\ S_{0} \,,
\qquad  S = \left(\begin{array}{cc}
	               1  & 0 \\
				   0  & -1 
\end{array} \right) \,.
\label{identity3}
\ee
We conclude that
\be
T_{0}(\lambda \pm {i\pi\over \mu}) = (-)^{N} S_{0}\ T_{0}(\lambda)\ S_{0} 
\,.
\ee 
In particular, we obtain the desired quasi-periodicity relations
\be
t(\lambda \pm {i\pi\over \mu}) = (-)^{N}\  t(\lambda) 
\label{quasi1}
\ee 
and
\be
B(\lambda \pm {i\pi\over \mu}) = (-)^{N+1}\  B(\lambda) \,.
\label{quasi2}
\ee
We shall make use of the latter relation when computing the parity of 
certain states.

\subsection{Charge conjugation}

The charge conjugation matrix $C$ is defined (see, e.g., 
\cite{baxter}) by \footnote{The charge conjugation matrix should not be 
confused with the element $C(\lambda)$ of the monodromy matrix!}
\be
C= \left(\begin{array}{cc}
	               0  & 1 \\
				   1  & 0 
\end{array} \right) \,,
\ee
since it interchanges the two-component spins ${1 \choose 0}$ and ${0 
\choose 1}$.  We denote by ${\cal C}$ the corresponding operator 
acting on the tensor product space $V^{\otimes N}$,
\be
{\cal C} =  C_{1} \cdots C_{N} \,.
\ee 
It has the properties ${\cal C} = {\cal C}^{-1} = {\cal C}^{\dagger}$,
and hence ${\cal C}\ {\cal C}^{\dagger}=1$.

The invariance of the $R$ matrix under charge conjugation
\be
C_{1}\ C_{2}\ R_{12}(\lambda)\ C_{1}\ C_{2} = R_{12}(\lambda)
\ee
implies that the $L$ operators obey
\be
C_{n}\ L_{0 n}(\lambda)\ C_{n} = C_{0}\ L_{0 n}(\lambda)\ C_{0} 
\,.
\ee
It follows that the monodromy matrix obeys
\be
{\cal C}\ T_{0}(\lambda)\ {\cal C} = C_{0}\ T_{0}(\lambda)\ C_{0} 
\,.
\ee 
In particular, we see that the transfer matrix is invariant under 
charge conjugation
\be
{\cal C}\ t(\lambda)\ {\cal C} = t(\lambda) \,,
\label{conj/first}
\ee
while the operator $B(\lambda)$ is mapped to $C(\lambda)$,
\be
{\cal C}\ B(\lambda)\ {\cal C} = C(\lambda) \,.
\label{conj/second}
\ee
Moreover, 
\be
{\cal C}\ \omega_{+} = \omega_{-} \,,
\label{conj/third}
\ee
where $\omega_{-}$ is the ferromagnetic vacuum vector with all spins 
down,
\be
\omega_{-} =  \underbrace{{0 \choose 1}
\otimes \cdots \otimes {0 \choose 1}}_{N} \,.
\ee

Since $C \sigma^{z} C = - \sigma^{z}$, the $z$-component of total spin 
changes sign under charge conjugation,
\be
{\cal C}\ S^{z}\ {\cal C} = -S^{z} \,.
\ee 
Therefore, a Bethe state can be an eigenstate of ${\cal C}$ only if 
$S^{z}=0$.  From Eq.  (\ref{spineigenvalue}), we see that this 
corresponds to $M=N/2$ with $N$ an even integer.

We conjecture that the eigenvectors (\ref{vec}) of the transfer matrix 
with $M=N/2$ are also eigenvectors of ${\cal C}$, with eigenvalues 
$(-)^{\nu}$, where
\be
\nu = {2 i \mu\over \pi} \sum_{\alpha=1}^{N/2} 
\lambda_{\alpha}  + {N\over 2} \qquad (\mbox{mod } 2 ) \,.
\label{conjecture}
\ee 
This conjecture is supported by explicit checks for $N=2$ and $N=4$, 
and it corresponds to the XXZ limit of a result 
\cite{baxter},\cite{jkm} for the XYZ chain.  Unfortunately, Baxter's 
$Q$-operator proof of the XYZ result, which relies on the 
quasi-double-periodicity of certain elliptic functions, does not 
survive the XXZ limit.  (The XYZ result is important, and so it is 
noteworthy that there does not appear to be a proof of it within the 
generalized algebraic Bethe Ansatz.) Certainly, Eqs.  
(\ref{conj/second}) and (\ref{conj/third}) alone do not seem to be 
sufficiently powerful to investigate this conjecture.

\section{Low-lying states}

We now examine some low-lying states of the critical XXZ chain within 
the framework of the string hypothesis \cite{takahashi/suzuki},
\cite{jkm}.
From the so-called root densities, we compute the parity and charge 
conjugation quantum numbers of the $S^{z}=0$ states, and we give a 
direct computation of the two-particle $S$ matrix.  In the analysis 
presented below, we find it convenient to work with the Bethe Ansatz 
Eqs.  (\ref{BAE}) in the form
\be 
e_{1}(\lambda_{\alpha}\,; \mu)^{N} = 
\prod_{\scriptstyle{\beta=1}\atop \scriptstyle{\beta \ne \alpha}}^M 
e_{2}(\lambda_{\alpha}-\lambda_{\beta}\,; \mu) \,, \quad \alpha = 1 
\,, \cdots \,, M \,,
\label{BAE/critical}
\ee
where
\be
e_{n}(\lambda\,; \mu) = 
{\sinh \mu \left( \lambda + {i n\over 2} \right) 
\over \sinh \mu \left( \lambda - {i n\over 2} \right) } \,.
\ee

\subsection{Ground state}

For simplicity, we henceforth restrict to the range
\be 
0 < \mu < {\pi\over 2} \,.
\ee 
The ground state then lies in the sector with $N$ even, and is 
characterized by $M=N/2$ real roots \cite{yang/yang}, 
\cite{takahashi/suzuki}.  We briefly review the procedure 
for determining the root density, which describes the distribution of 
roots in the thermodynamic ($N \rightarrow \infty$) limit.  Taking the 
logarithm of the Bethe Ansatz Eqs.  (\ref{BAE/critical}), we obtain
\be 
h (\lambda_\alpha) = J_\alpha \,,  
\qquad \alpha = 1 \,, \cdots \,, M \,,
\label{BAlog}
\ee 
where the so-called counting function $h(\lambda)$ is given by
\be 
h(\lambda) = {1\over 2\pi} \left\{  N q_1(\lambda\,; \mu) 
- \sum_{\beta=1}^{M}  q_{2}(\lambda - \lambda_{\beta}\,; \mu) 
\right\} \,, 
\label{counting} 
\ee 
$q_n (\lambda\,; \mu)$ is an odd function of $\lambda$ defined by
\be
q_n (\lambda\,; \mu) = \pi + i\log e_n(\lambda\,; \mu) \,,
\label{qr}
\ee
and $\{ J_\alpha \}$ are integers or half-integers lying in a certain 
range which serve as ``quantum numbers'' of the Bethe Ansatz states.  
The root density $\sigma(\lambda)$ is defined by
\be
\sigma(\lambda) = {1\over N} {d \over d \lambda} h(\lambda) \,,
\label{sigma}
\ee 
so that the number of $\lambda_{\alpha}$ in the interval $[ \lambda 
\,, \lambda + d\lambda ]$ is $N \sigma(\lambda) d\lambda$.
Passing from the sum in $h(\lambda)$ to an integral, we obtain a
linear integral equation for the root density
\be
\sigma(\lambda) = a_{1}(\lambda \,; \mu) 
- \int_{-\infty}^{\infty} d\lambda' \sigma(\lambda')\ 
a_{2}(\lambda - \lambda'\,; \mu) \,,
\ee 
where 
\be
a_n(\lambda\,; \mu) = {1\over 2\pi} {d \over d\lambda} q_n (\lambda\,; \mu)
= {\mu \over \pi} 
{\sin (n \mu)\over \cosh(2 \mu \lambda) - \cos (n \mu)} 
\,.
\ee 
Solving by Fourier transforms \footnote{Our 
conventions in the critical regime are
\be
\hat f(\omega) \equiv \int_{-\infty}^\infty e^{i \omega \lambda}\ 
f(\lambda)\ d\lambda \,, \qquad\qquad
f(\lambda) = {1\over 2\pi} \int_{-\infty}^\infty e^{-i \omega \lambda}\ 
\hat f(\omega)\ d\omega \,, \non 
\ee 
and we use $*$ to denote the convolution
\be
\left( f * g \right) (\lambda) = \int_{-\infty}^\infty 
f(\lambda - \lambda')\ g(\lambda')\ d\lambda' \,. \non 
\ee  
} using
\be
\hat a_{n}(\omega\,; \mu) = {\sinh \left( ({\pi \over \mu}  - n) 
{\omega \over 2}) \right) \over
\sinh \left( {\pi \over \mu} {\omega \over 2} \right)} \,,
\qquad 0 < n < {2\pi\over \mu} \,, 
\ee
we conclude that the root density for the ground state is given by
\be
\sigma(\lambda) = s(\lambda) = {1\over 2\pi} 
\int_{-\infty}^\infty d\omega\ e^{-i \omega \lambda}\ \hat s(\omega)
 = {1\over 2 \cosh \left( \pi \lambda \right)} \,,
\label{density/ground}
\ee 
where
\be
\hat s(\omega) = {\hat a_{1}(\omega\,; \mu)
\over 1 + \hat a_{2}(\omega\,; \mu)} = 
{1\over 2 \cosh \left( {\omega\over 2} \right)}
\,.
\ee 

We verify the consistency of this procedure by computing the value of 
$M$ from the root density:
\be
M = \sum_{\alpha=1}^{M} 1 
= N \int_{-\infty}^{\infty}d\lambda\ \sigma(\lambda) 
= N \hat \sigma (0) = {N\over 2} \,,
\label{consistency}
\ee
and hence, the state indeed has $S^{z}=0$. The energy and momentum are
\be
E_{gr} &=& -{\pi \sin \mu \over \mu } \sum_{\alpha=1}^{N/2}
a_{1}(\lambda_{\alpha}) =  
-{\pi \sin \mu \over \mu } 
N \int_{-\infty}^{\infty}d\lambda\ s(\lambda)\ a_{1}(\lambda) \non \\
P_{gr} &=& - \sum_{\alpha=1}^{N/2} \left[ 
q_{1}(\lambda_{\alpha}) - \pi \right] = {\pi N\over 2} 
\quad (\mbox{mod } 2 \pi)\,.
\ee 

The ground state is nondegenerate \cite{yang/yang}, and therefore, as 
argued at the end of Section 2.1 , this state must be a parity 
eigenstate with eigenvalue $(-)^{N/2}$.  We now give an alternative 
derivation, in order to illustrate the line of argument which we shall 
use to compute the parity of excited states.  Denoting the ground 
state by $|v \rangle $, we have
\be
|v \rangle &=& \prod_{\alpha=1}^{M} B(\lambda_{\alpha})\ \omega_{+} 
= \exp \left( \sum_{\alpha=1}^{M} \log B(\lambda_{\alpha}) \right)\ 
\omega_{+} \non  \\
&=& \exp \left( N \int_{-\infty}^{\infty}d\lambda\ \sigma(\lambda) 
\log B(\lambda) \right)\ \omega_{+} \,.
\label{one}
\ee
Moreover, using (\ref{parity/second}), we obtain
\be
\Pi\ |v \rangle &=& (-)^{M} \prod_{\alpha=1}^{M} B(-\lambda_{\alpha})\ 
\omega_{+} \non  \\
&=& (-)^{M} \exp \left( N \int_{-\infty}^{\infty}d\lambda\ \sigma(\lambda) 
\log B(-\lambda) \right)\ \omega_{+} \non  \\
&=& (-)^{M} \exp \left( N \int_{-\infty}^{\infty}d\lambda\ \sigma(-\lambda) 
\log B(\lambda) \right)\ \omega_{+} \,,
\label{two}
\ee
where in passing to the last line we have made the change of variables 
$\lambda \rightarrow -\lambda$.  Finally, comparing Eqs.  (\ref{one}) 
and (\ref{two}), and using the fact that the root density 
(\ref{density/ground}) is an even function $\sigma(-\lambda) = 
\sigma(\lambda)$, we conclude that

\be
\Pi\ |v \rangle = (-)^{N/2} |v \rangle \,.
\label{three}
\ee 

The nondegeneracy of the ground state also implies that this state is 
a charge conjugation eigenstate. The formula (\ref{conjecture}) 
implies that the corresponding eigenvalue is also $(-)^{N/2}$, since
\be
\sum_{\alpha=1}^{N/2} \lambda_{\alpha}
= N \int_{-\infty}^{\infty}d\lambda\ \sigma(\lambda)\ \lambda = 0 \,,
\ee
where again we have made use of the fact that the root density is an 
even function.

\subsection{Two-particle excited states}

We consider now two-particle excited states, again with $N$ even.  On 
the basis of the results for the isotropic XXX chain 
\cite{faddeev/takhtajan}, one expects that the particles have $S^{z} = 
\pm 1/2$.  Hence, one expects four two-particle excited states: two 
states with $S^{z} = \pm 1$, and two $S^{z} = 0$ states which are 
distinguished by their parity and charge conjugation quantum numbers.  
Surprisingly, the Bethe Ansatz does not seem to give the $S^{z}= \pm 
1$ states.  Indeed, we find that the Bethe Ansatz state with two holes 
and no strings has {\it fractional} spin (see Eq.  (\ref{fracspin}) 
below), which only for $\mu \rightarrow 0$ (i.e., the isotropic limit) 
is equal to 1.  We do find two $S^{z} = 0$ states 
\cite{jkm},\cite{woynarovich} which indeed are distinguished by their 
parity and charge conjugation quantum numbers.

\vspace{.2in}
\noindent
{\it (0) Two holes}
\vspace{.2in}

We consider first the two-hole state with only real roots.  We label 
the holes by the integers or half-integers $\{ \tilde J_\alpha \} \,,\
\alpha = 1 \,, 2$.  The corresponding hole rapidities $\{ 
\tilde\lambda_{\alpha} \}$ are defined by 
\be 
h (\tilde\lambda_{\alpha}) = \tilde J_{\alpha} \,, 
\qquad \alpha = 1 \,, 2 \,, 
\ee 
where the counting function is given by Eq. (\ref{counting}).
We compute the density for this state by the same procedure 
used for the ground state, except that now in passing from sums to 
integrals we must take into account the presence of holes, i.e.,
\be 
{1\over N}\sum_{\alpha=1}^{M} g(\lambda_\alpha)
= \int_{-\infty}^{\infty} d\lambda \  \sigma(\lambda)\ g(\lambda) 
- {1\over N}\sum_{\alpha=1}^{2} g(\tilde\lambda_\alpha) \,,
\label{euler} 
\ee
for any function $g(\lambda)$. We find
\be
\sigma_{(0)}(\lambda) = s(\lambda) + {1\over N} 
\sum_{\alpha=1}^{2} J(\lambda - \tilde\lambda_\alpha) 
\,, \label{density/(0)}
\ee
where 
\be
\hat J(\omega) = {\hat a_{2}(\omega\,; \mu)\over 1 
+ \hat a_{2}(\omega\,; \mu)} =
{\sinh \left( ({\pi \over \mu}  - 2) 
{\omega \over 2}) \right) \over
2 \sinh \left( ({\pi \over \mu} - 1){\omega \over 2} \right)
\cosh \left( {\omega \over 2} \right)} \,.
\ee

The energy and momentum are given by
\be
E = E_{gr} + {\pi \sin \mu \over \mu } \sum_{\alpha=1}^{2} 
s(\tilde\lambda_{\alpha}) \,, \qquad 
P = P_{gr} +  \sum_{\alpha=1}^{2} p(\tilde\lambda_{\alpha}) \,,
\label{energy/mom/exc/crit}
\ee 
where the hole momentum $p(\lambda) \equiv \left( q_{1} -  J * q_{1}
\right)(\lambda) $ satisfies
\be
{d\over d\lambda} p(\lambda) = 2 \pi s(\lambda) \,. 
\label{derivmom}
\ee 

We find by a generalization of the computation (\ref{consistency}) 
that this two-particle state has the fractional spin
\be
S^{z} = {\pi \over \pi - \mu} \,.
\label{fracspin}
\ee
A similar result has been obtained by thermodynamic 
arguments \cite{babujian/tsvelick}, \cite{kirillov/reshetikhin}. This 
result is puzzling for both mathematical and physical reasons: the formula 
(\ref{spineigenvalue}) for the spin eigenvalue implies that $S^{z}$ 
should be an integer; and as already mentioned, the results for the 
isotropic XXX chain suggest that this state should have $S^{z} = 1$, 
corresponding to two spin $1/2$ particles with spins ``up''.  
% The authors of Ref.  \cite{kirillov/reshetikhin} define a ``spin 
% renormalization factor'', such that the ``renormalized spin'' of the 
% particles is $1/2$. 
A possible resolution is to take in Eq. (\ref{euler}) {\it finite} 
integration limits $\pm \Lambda$ such that
\be
\int_{-\Lambda}^{\Lambda} d \lambda \ \sigma(\lambda) = {1\over 2} + 
{1\over N} \,, \non
\ee
which would ensure $S^{z}=1$.

We do not attempt to compute the parity of this state, since the value 
of $M$ is not an integer.  Since this state has $S^{z} \ne 0$, it 
cannot be an eigenstate of ${\cal C}$.  Indeed, by acting on this 
state with the charge conjugation operator ${\cal C}$, one obtains the 
$S^{z} = -{\pi \over \pi - \mu}$ state.

\vspace{.2in}
\noindent
{\it (a) Two holes and one 2-string}
\vspace{.2in}

We consider the two-hole state with one string of length 2 (i.e., a 
pair of roots of the form $\lambda_{0} \pm {i\over 2}$, with 
$\lambda_{0}$ real) and all other roots real. \footnote{This ``positive 
parity'' 2-string satisfies the Takahashi-Suzuki 
\cite{takahashi/suzuki} conditions for $0 \le \mu < \pi$.} The Bethe 
Ansatz Eqs.  (\ref{BAE/critical}) imply
\be
e_{1}(\lambda_{\alpha}\,; \mu)^{N} & = &
e_{1}(\lambda_{\alpha}-\lambda_{0}\,; \mu) \  
e_{3}(\lambda_{\alpha}-\lambda_{0}\,; \mu) 
\prod_{\scriptstyle{\beta=1}\atop \scriptstyle{\beta \ne \alpha}}^{M_{1}^{+}} 
e_{2}(\lambda_{\alpha}-\lambda_{\beta}\,; \mu) \,,  \non  \\ 
& & \qquad  \alpha = 1 \,, \cdots \,, M_{1}^{+} \,, \label{BAEa1} \\ 
e_{2}(\lambda_{0}\,; \mu)^{N} & = &
\prod_{\beta=1}^{M_{1}^{+}}
e_{1}(\lambda_{0}-\lambda_{\beta}\,; \mu)  \ 
e_{3}(\lambda_{0}-\lambda_{\beta}\,; \mu) \,,
\label{BAEa2}
\ee
where $M_{1}^{+}$ is the number of real roots, i.e., $M = M_{1}^{+} + 2$.
The counting function is therefore now given by
\be 
h(\lambda) &= & {1\over 2\pi} \Bigl\{ N q_1(\lambda\,; \mu) 
- \sum_{\beta=1}^{M_{1}^{+}}  q_{2}(\lambda - \lambda_{\beta}\,; \mu)  
\non  \\ 
& & - \left[ q_{1}(\lambda - \lambda_{0}\,; \mu) 
+ q_{3}(\lambda - \lambda_{0}\,; \mu) \right] \Bigr\}
\,.
\ee 
Proceeding as in case (0), we find
\be
\sigma_{(a)}(\lambda) = s(\lambda) + {1\over N} \left[
\sum_{\alpha=1}^{2} J(\lambda - \tilde\lambda_\alpha) 
- a_{1}(\lambda - \lambda_{0}\,; \mu') \right]
\,, \label{density/(a)}
\ee
where $\mu'$ is the ``renormalized'' anisotropy parameter given by
\be
{\pi \over \mu'} = {\pi \over \mu} - 1   \,, \non 
\ee
that is,
\be
\mu' = {\pi \mu \over \pi - \mu } \,.
\label{renormalized}
\ee 
In order to cast the last term in the expression for 
$\sigma_{(a)}(\lambda)$ as the function $a_{1}$, it is 
essential to use $\mu'$ rather than $\mu$.

The center of the 2-string, $\lambda_{0}$, can be determined from
Eq. (\ref{BAEa2}). Taking the logarithm and passing from a sum to 
an integral, one obtains the condition
\be
\sum_{\alpha=1}^{2} q_1(\lambda_{0} - \tilde\lambda_{\alpha} \,; \mu') 
= 0 \,, 
\label{centercondition}
\ee
which has the solution
\be
\lambda_{0} = {1\over 2} \left( \tilde\lambda_{1} + 
\tilde\lambda_{2} \right) \,. 
\label{center}
\ee 

We now investigate the quantum numbers of this state:
\begin{itemize}
	
	\item We verify by a generalization of the 
computation (\ref{consistency}) that this state has $S^{z}=0$. 
The energy and momentum are again given by Eq.  
(\ref{energy/mom/exc/crit}).

    \item Since the density $\sigma_{(a)}(\lambda) $ is not an even 
function of $\lambda$ for generic values of $\{ \tilde\lambda_{\alpha} 
\}$ , a generalization of the argument (\ref{one}) - (\ref{three}) 
implies that this state is not a parity eigenstate.  However, in the 
``rest frame'' \cite{korepin}
\be
\tilde\lambda_{1} + \tilde\lambda_{2} = 0 \,,
\label{rest}
\ee
the density is an even function, and therefore the state is a parity 
eigenstate, with parity $(-)^{N/2}$. We note that in the rest frame, 
the momentum is $P = 0 \mbox{  or  } \pi \quad (\mbox{mod } 2 \pi)$, 
which is consistent with the fact $\{ \Pi \,, P \} = 0$.

    \item According to the conjecture of Section 2.3, this state is an 
eigenstate of charge conjugation for all values of 
$\{ \tilde\lambda_{\alpha} \}$, with eigenvalue given by the formula 
(\ref{conjecture}). Remarkably, the first term of that formula gives a 
vanishing contribution:
\be
\sum_{\alpha=1}^{N/2}\lambda_{\alpha} &=& 
\bigl( \lambda_{0} + {i\over 2} \bigr) 
+ \bigl( \lambda_{0} - {i\over 2} \bigr)
+ \sum_{\alpha=1}^{M_{1}^{+}}\lambda_{\alpha} \non  \\
&=& 2 \lambda_{0} + N \int_{-\infty}^{\infty}d\lambda\ 
\sigma(\lambda)\ \lambda 
- \sum_{\alpha=1}^{2} \tilde\lambda_{\alpha} \non  \\
&=& N \int_{-\infty}^{\infty}d\lambda\ 
\sigma(\lambda)\ \lambda  \non  \\
&=& \sum_{\alpha=1}^{2} \int_{-\infty}^{\infty}d\lambda\ 
J(\lambda - \tilde\lambda_\alpha) \ \lambda 
- \int_{-\infty}^{\infty}d\lambda\ 
a_{1}(\lambda - \lambda_{0}\,; \mu') \ \lambda \non  \\
&=& \hat J(0) \sum_{\alpha=1}^{2} \tilde\lambda_{\alpha} 
- \hat a_{1}(0\,; \mu')\ \lambda_{0} \non  \\
&=& 0 \,,
\label{remarkable}
\ee 
where we have used the relation (\ref{center}) twice, and also the 
fact that $\hat J(\omega)$ and $\hat a_{1}(\omega\,; \mu')$ are even 
functions of $\omega$, and therefore have a vanishing first derivative 
at $\omega=0$.  We conclude that the charge conjugation 
eigenvalue for this state is also $(-)^{N/2}$.

\end{itemize}

\vspace{.2in}
\noindent
{\it (b) Two holes and one negative-parity 1-string}
\vspace{.2in}

We now consider the two-hole state with one ``negative-parity'' string 
of length 1 (i.e., a root of the form $\lambda_{0} + {i \pi \over 2 
\mu}$, with $\lambda_{0}$ real) \cite{takahashi/suzuki} and all other 
roots real.  The Bethe Ansatz Eqs.  (\ref{BAE/critical}) imply
\be
e_{1}(\lambda_{\alpha}\,; \mu)^{N} & = &
g_{2}(\lambda_{\alpha}-\lambda_{0}\,; \mu)  
\prod_{\scriptstyle{\beta=1}\atop 
\scriptstyle{\beta \ne \alpha}}^{M_{1}^{+}} 
e_{2}(\lambda_{\alpha}-\lambda_{\beta}\,; \mu) \,,  \non  \\ 
& & \qquad  \alpha = 1 \,, \cdots \,, M_{1}^{+} \,, \label{BAEb1} \\ 
g_{1}(\lambda_{0}\,; \mu)^{N} & = &
\prod_{\beta=1}^{M_{1}^{+}}
g_{2}(\lambda_{0}-\lambda_{\beta}\,; \mu) \,,
\label{BAEb2}
\ee
where
\be
g_{n}(\lambda \,; \mu) = e_{n}(\lambda \pm {i \pi \over 2 \mu}\,; \mu)
= {\cosh \mu \left( \lambda + {i n\over 2} \right) 
\over \cosh \mu \left( \lambda - {i n\over 2} \right) } \,,
\ee 
and $M = M_{1}^{+} + 1$. The counting function is now given by
\be 
h(\lambda) = {1\over 2\pi} \left\{  N q_1(\lambda\,; \mu) 
- r_{2}(\lambda - \lambda_{0}\,; \mu) 
- \sum_{\beta=1}^{M_{1}^{+}}  q_{2}(\lambda - \lambda_{\beta}\,; \mu) 
\right\} \,, 
\ee 
where $r_n (\lambda\,; \mu)$ is an odd function of $\lambda$ defined by
\be
r_n (\lambda\,; \mu) = i\log g_n(\lambda\,; \mu) \,.
\ee
Proceeding as before, we obtain the density
\be
\sigma_{(b)}(\lambda) = s(\lambda) + {1\over N} \left[
\sum_{\alpha=1}^{2} J(\lambda - \tilde\lambda_\alpha) 
- b_{1}(\lambda - \lambda_{0}\,; \mu') \right]
\,, \label{density/(b)}
\ee
where
\be
b_n(\lambda\,; \mu) &=& {1\over 2\pi} {d \over d\lambda} r_n (\lambda\,; \mu)
= -{\mu \over \pi} 
{\sin (n \mu)\over \cosh(2 \mu \lambda) + \cos (n \mu)} \,, \non  \\
\hat b_{n}(\omega\,; \mu) &=& 
- {\sinh \left( {n \omega \over 2} \right) \over
\sinh \left( {\pi \over \mu} {\omega \over 2} \right)} \,, \qquad 
0 < n < {\pi\over \mu} \,, 
\ee
and $\mu'$ is given by Eq.  
(\ref{renormalized}).  We determine the center of the negative-parity 
1-string, $\lambda_{0}$, from Eq.  (\ref{BAEb2}) by the same procedure 
used in case (a), and we obtain the same result (\ref{center}).

Let us examine the quantum numbers of this state:
\begin{itemize}
	
	\item  We verify by a generalization of the computation 
(\ref{consistency}) that this state also has $S^{z}=0$. The energy 
and momentum are again given by Eq. (\ref{energy/mom/exc/crit}).

    \item  As in case (a), this state is not a parity eigenstate
for generic values of $\{ \tilde\lambda_{\alpha} \}$. Let us now 
restrict to the rest frame (\ref{rest}). The Bethe vector is 
then given by
\be
|v \rangle = B({i \pi \over 2 \mu}) \ 
\prod_{\alpha=1}^{{N\over 2}-1} B(\lambda_{\alpha}) \ \omega_{+} \,.
\ee 
Acting with the parity operator using Eq.  (\ref{parity/second}), we 
obtain
\be
\Pi\ |v \rangle &=& (-)^{N/2} B(-{i \pi \over 2 \mu}) \ 
\prod_{\alpha=1}^{{N\over 2}-1} B(-\lambda_{\alpha}) \ \omega_{+} 
\non  \\ 
 &=& - (-)^{N/2} B({i \pi \over 2 \mu}) \ 
\prod_{\alpha=1}^{{N\over 2}-1} B(-\lambda_{\alpha}) \ \omega_{+} 
\non  \\ 
&=& - (-)^{N/2} |v \rangle \,.
\ee 
In passing to the second line, we have used the quasi-periodicity 
property (\ref{quasi2}), and to arrive at the last line we use the 
fact that (in the rest frame) the density is an even function of 
$\lambda$ .  Thus, the state has parity $-(-)^{N/2}$. The appellation 
``negative parity'' for this string is quite apt!

    \item Unlike case (a), here the first term of the formula
(\ref{conjecture}) for the charge conjugation eigenvalue does give a 
contribution, namely, $-1$.  (The computation closely parallels the 
one (\ref{remarkable}) for case (a).) We conclude that the charge 
conjugation eigenvalue for this state is also $-(-)^{N/2}$.

\end{itemize}

We recall that a Boson-antiBoson state with a symmetric 
(antisymmetric) wavefunction has positive (negative) $\Pi$ and ${\cal 
C}$, while for a Fermion-antiFermion state the opposite is true.  
(See, e.g., Refs.  \cite{bjorken/drell}, \cite{klassen/melzer}.) 
Evidently the statistics of the XXZ excitations vary with the 
value of $N$.

\subsection{$S$ matrix}

We define the $S$ matrix $S(\tilde\lambda_1, \tilde\lambda_2)$ by the 
momentum quantization condition \cite{korepin},\cite{andrei/destri}
\be
\left(e^{i p(\tilde\lambda_1) N}\
S(\tilde\lambda_1, \tilde\lambda_2) - 1 \right) 
| \tilde\lambda_1, \tilde\lambda_2 \rangle = 0 \,, 
\label{quantization} 
\ee
where $\tilde\lambda_1$, $\tilde\lambda_2$ are the 
hole rapidities, and $p(\lambda)$ is the hole momentum.
Combining the definition (\ref{sigma}) of the root density with the 
relation (\ref{derivmom}) for the hole momentum, we immediately obtain 
the identity
\be 
{d\over d\lambda} p(\lambda) 
+ 2\pi \left( \sigma(\lambda) - s(\lambda) \right) = 
{2\pi\over N} {d\over d\lambda} h(\lambda) \,.
\ee
Integrating this equation from $-\infty$ to $\tilde\lambda_1$, noting 
that $h(\tilde\lambda_1)=\tilde J_{1}$, and then comparing with Eq.  
(\ref{quantization}), we see that the $S$ matrix eigenvalues are given 
(up to a rapidity-independent phase factor) by
\be
S_{(j)} \sim   
\exp \left\{ i 2\pi N \int_{-\infty}^{\tilde\lambda_{1}}
\left( \sigma_{(j)}(\lambda) - s(\lambda) \right) d\lambda
\right\}
\,, \qquad j = 0 \,, a \,, b \,, 
\label{sim}
\ee 
where $S_{(0)}$, $S_{(a)}$ and $S_{(b)}$ are the eigenvalues of the 
$S$ matrix corresponding to states (0), (a) and (b), respectively.

Recalling the expression (\ref{density/(0)}) for 
the density $\sigma_{(0)}(\lambda)$, we obtain
\be
S_{(0)} & \sim  & \exp \left\{ i 2\pi \sum_{\alpha=1}^{2} 
\int_{-\infty}^{\tilde\lambda_{1}}
J(\lambda - \tilde\lambda_\alpha) \ d\lambda \right\} \non  \\
&=& \exp \left\{ \int_{0}^{\infty} {d\omega\over \omega}
{\sinh \left( ({\pi \over 2\mu'}  - {1\over 2}) \omega \right) 
\sinh \left( i \omega \tilde\lambda \right) \over
\sinh \left( {\pi \omega\over 2\mu'}  \right)
\cosh \left( {\omega \over 2} \right)} \right\}  \non  \\
&=& \prod_{n=0}^{\infty} \Bigl\{ 
{\Gamma \left[ \left( 1 + {\pi\over \mu'} n - i \tilde \lambda 
\right)/2 \right]\over
\Gamma \left[ \left( 1 + {\pi\over \mu'} n + i \tilde \lambda 
\right)/2 \right]}
{\Gamma \left[ \left( 2 + {\pi\over \mu'} n + i \tilde \lambda 
\right)/2 \right]\over
\Gamma \left[ \left( 2 + {\pi\over \mu'} n - i \tilde \lambda 
\right)/2 \right]} \non  \\
& & \times 
{\Gamma \left[ \left( {\pi\over \mu'} ( n + 1 ) + i \tilde \lambda 
\right)/2 \right]\over
\Gamma \left[ \left( {\pi\over \mu'} ( n + 1 ) - i \tilde \lambda 
\right)/2 \right]}
{\Gamma \left[ \left( 1 + {\pi\over \mu'} ( n + 1 )  - i \tilde \lambda 
\right)/2 \right]\over
\Gamma \left[ \left( 1 + {\pi\over \mu'} ( n + 1 )  + i \tilde \lambda 
\right)/2 \right]} \Bigr\} \,,
\label{common}
\ee 
where $\tilde\lambda = \tilde\lambda_{1} - \tilde\lambda_{2}$.
Moreover, recalling the expressions (\ref{density/(a)}) and 
(\ref{density/(b)}) for the densities $\sigma_{(a)}(\lambda)$ and 
$\sigma_{(b)}(\lambda)$, we obtain
\be
{S_{(a)}\over S_{(0)}} &=& \exp \left\{ - i 2\pi 
\int_{-\infty}^{\tilde\lambda_{1}}
a_{1}(\lambda - \lambda_{0}\,; \mu') \ d\lambda \right\} 
= e_{1}({\tilde\lambda \over 2}\,; \mu') \,, \non \\
{S_{(b)}\over S_{(0)}} &=& \exp \left\{ - i 2\pi 
\int_{-\infty}^{\tilde\lambda_{1}}
b_{1}(\lambda - \lambda_{0}\,; \mu') \ d\lambda \right\} 
= g_{1}({\tilde\lambda \over 2}\,; \mu') \,,
\ee 
where again we have used (\ref{center}).  We conclude that the $S$ 
matrix for the critical XXZ chain is given by
\be
S_{(a)} = S_{(0)}\  
{\sinh \left( \mu' ( \tilde\lambda + i )/2 \right) \over 
 \sinh \left( \mu' ( \tilde\lambda - i )/2 \right) } \,, \qquad 
S_{(b)} = S_{(0)}\  
{\cosh \left( \mu' ( \tilde\lambda + i )/2 \right) \over 
 \cosh \left( \mu' ( \tilde\lambda - i )/2 \right) } \,,
\ee 
where $S_{(0)}$ is given by Eq. (\ref{common}), and $\mu'$ is given by
Eq. (\ref{renormalized}).
This coincides with the $S$ matrix of sine-Gordon/massive Thirring 
model \cite{korepin},\cite{karowski},\cite{zamolodchikov}, provided we 
identify the sine-Gordon coupling constant $\beta^{2}$ as
\be
\beta^{2} = 8 \left( \pi - \mu \right) \,.
\ee 
This result has been obtained for the XXZ chain previously, although 
by less direct means, in Refs.  \cite{babujian/tsvelick}, 
\cite{kirillov/reshetikhin}.  Note that the regime $0 < \mu <
{\pi\over 2}$ in which we work corresponds to the ``repulsive'' 
regime $4\pi < \beta^{2} < 8 \pi$.

\section{Outlook}

A number of issues remain to be explored.  It would be interesting to 
find a proof (or counterexample!) of the formula (\ref{conjecture}) 
for the charge conjugation eigenvalues of the Bethe Ansatz states, and 
to better understand states in the critical regime with $S^{z} \ne 0$.  
It may be worthwhile to investigate discrete symmetries in integrable 
chains constructed with higher-rank $R$ matrices, 
\cite{bazhanov},\cite{jimbo} such as $A_{{\mathcal N}-1}^{(1)}$ with 
${\mathcal N} > 2$.  Since these $R$ matrices are not parity 
invariant, neither are the corresponding Hamiltonians.  However, the 
$R$ matrices do have PT symmetry, which may lead to a useful symmetry 
on the space of states.  Moreover, we have not discussed here the 
interesting case of the noncritical ($\Delta > 1$) regime \cite{unpub}.

\section*{Acknowledgments}

We thank F. Essler and V. Korepin for valuable discussions.
This work was supported in part by the National Science Foundation 
under Grant PHY-9870101.

% We dedicate this paper to Professor Jim McGuire on the occasion of his 
% sixty-fifth birthday.  The interaction with our near neighbor has been 
% for us one of the unexpected pleasures of entering the field of 
% integrable models.

\end{document}